\documentclass[11pt]{article}

\usepackage[margin=1in]{geometry}
\usepackage{amsmath,amssymb,amsthm}
\usepackage{graphicx}
\usepackage{booktabs}
\usepackage[dvipsnames]{xcolor}
\usepackage{hyperref}
\usepackage{microtype}
\usepackage{enumitem}
\usepackage{caption}

\newtheorem{proposition}{Proposition}

\hypersetup{
  colorlinks=true,
  linkcolor=blue!60!black,
  citecolor=blue!60!black,
  urlcolor=blue!60!black
}

\newcommand{\dd}{\mathrm{d}}
\newcommand{\bR}{\mathbb{R}}
\newcommand{\range}{\mathrm{range}}
\newcommand{\kers}{\mathrm{ker}}

\title{\textbf{Geometric Realism Without Angular Resolution}\\[4pt]
Structural Classification of Multilayer Kubelka--Munk Theory\\
within Radiative Transport}

\author{Claude Zeller\\[2pt]
\textit{Claude Zeller Consulting LLC, Tillamook, Oregon}\\
\texttt{czeller@ieee.org}}

\date{}

\begin{document}
\maketitle

\begin{abstract}
Kubelka--Munk (KM) theory provides a two-flux description of radiative
transport in layered scattering and absorbing media.  Despite its wide
use in the coatings, paper, paint, and textile industries, the theory
has often been regarded as a phenomenological model whose connection to
the full radiative transfer equation (RTE) remains unclear.  Under the
standard steady-state, plane-parallel, azimuthally symmetric assumptions,
we show that multilayer KM theory is exactly a rank-2 Galerkin projection
of the RTE onto hemispherical basis functions.  The projection
is idempotent with an infinite-dimensional kernel, and its rank is
preserved under multilayer composition---so no amount of layer stacking
can recover angular information discarded by the projection.  We derive
the KM coefficients as hemispherical moments of the transport operator
and compute the projection error for representative scattering media
($g$ from~0 to~0.85), finding that the reduced optical thickness
$\tau^*=\tau(1-g)$ governs KM accuracy.  The projection-error framework
explains the well-documented accuracy of compositional multilayer models
in printed media and shows where higher-order methods become necessary.  The result places KM theory on
rigorous footing as a legitimate---if low-resolution---transport
approximation rather than an ad hoc phenomenology.
\end{abstract}

\section{Introduction}

Two-flux models collapse the full angular distribution of radiance into
a pair of scalar fluxes: one going forward, one going backward.  It is
about the crudest simplification imaginable, and it has been
extraordinarily successful.  The Kubelka--Munk (KM) model~\cite{KM1931,
Kubelka1948}, introduced in 1931, is the most widely used instance.  It
describes a homogeneous scattering-absorbing slab using two
parameters---an absorption coefficient~$K$ and a scattering
coefficient~$S$---and yields closed-form expressions for reflectance and
transmittance.  For decades it has been the workhorse of color science in
paints, paper, textiles, and coatings.

The simplicity that makes KM theory useful also makes it suspect.  The
model cannot distinguish an isotropic scatterer from one with a sharp
forward peak.  It treats every photon within a hemisphere as
interchangeable.  Many authors have regarded it as ``merely
phenomenological'' (see, e.g., the discussion in~\cite{Vargas1997}) and have proposed corrections---Saunderson surface
corrections~\cite{Saunderson1942}, path-length multipliers, modified
scattering coefficients---that try to put back what the two-flux
reduction threw away.

The connection between KM theory and the radiative transfer equation is
not itself new.  Mudgett and Richards~\cite{Mudgett1971} derived KM from
the RTE by hemispherical averaging and established that $K=2\sigma_a$.
Star et al.~\cite{Star1988} and van Gemert and Star~\cite{vanGemert1987}
refined the treatment for anisotropic scattering, identifying the
backscatter fraction $\bar{p}_{-+}$ as the bridge between transport and
KM scattering coefficients.  These derivations are correct and this
paper relies on them.  What they did not do is identify the
hemispherical average as an orthogonal Galerkin projection with a
specific rank, kernel, and idempotence structure.  That step is what
this paper adds, and it yields three results that do not follow from the
earlier work: (i)~the infinite-dimensional kernel $\kers(\mathcal{P})$,
which characterizes exactly what KM theory cannot represent; (ii)~a
rank-preservation theorem explaining why multilayer composition cannot
recover the discarded information; and (iii)~a projection-error
criterion $\varepsilon$ that quantitatively bounds KM accuracy as a
function of the scattering parameters.

A substantial body of work by H\'ebert, Hersch, and
collaborators~\cite{Hebert2006, Hebert2007, Mazauric2014, Hebert2015,
Simonot2016} has developed a powerful compositional framework for
predicting the reflectance and transmittance of multilayer specimens
using transfer matrices.  Each optical element---a scattering layer, an
ink film, a refractive interface---is characterized by its reflectances
and transmittances.  Stacking two elements produces a composite element
whose properties follow from summing the infinite geometric series of
internal bounces in closed form.  They showed that Kubelka's layering
model, the Saunderson correction, and the Williams--Clapper model are
all special cases of a single compositional formalism, and that the
two-flux matrix approach extends naturally to four-flux and multiflux
models when one needs to distinguish collimated from diffuse light.  The
framework has been validated to high accuracy in printed paper,
ink-on-substrate systems, and stacked transparencies---application
domains where KM theory is known to work well.

The H\'ebert--Hersch framework operates at the level of the KM
quantities themselves: the phase function $p(\mu,\mu')$, the asymmetry
parameter~$g$, and the distinction between forward-peaked and isotropic
scattering do not appear in the compositional algebra, nor do they need
to in those application domains.  This raises a question that is not
immediately obvious: why should a rank-2 model---one that cannot
distinguish $g=0.90$ from $g=0.99$---succeed so consistently in
predicting color-accurate reflectance for real printed systems?
The present paper addresses the logically prior question that the
compositional approach takes as given:
what, exactly, is the relationship between the full angular transport
problem and the two-flux description?  Where does the phase function
enter?  Where does it get lost?  And why does the transfer-matrix
algebra work so well?

A useful way to situate Kubelka--Munk theory within the broader family
of transport approximations is to recognize that many classical models
arise from different projections of the same radiative transfer operator
onto different angular subspaces.  In the present formulation, the
two-flux Kubelka--Munk equations correspond to a projection onto the
hemispherical subspace spanned by the indicator functions
$\{\chi^+,\chi^-\}$, which retain only the directional sign of the
radiance (forward versus backward) while discarding all finer angular
structure.  By contrast, the diffusion or $P_1$ approximation projects
onto the Legendre subspace $\{1,\mu\}$, preserving the first angular
moment and therefore the angular gradient of the intensity.  Discrete
ordinates methods ($S_N$) can likewise be interpreted as projections onto
a set of directional delta functions representing selected propagation
angles.  From this viewpoint these methods are not competing theories but
different low-dimensional reductions of the same infinite-dimensional
transport problem, each retaining different angular information.
Kubelka--Munk sits at the extreme end of this hierarchy: it preserves
geometric directionality (forward versus backward flux) while discarding
all internal angular resolution within each hemisphere.  Understanding KM
theory in this projection framework clarifies both its remarkable
practical success in multiply scattering media and its intrinsic
limitations in strongly anisotropic or optically thin regimes.

We show that the KM equations arise exactly as a rank-2 Galerkin
projection of the full radiative transfer equation onto the subspace
spanned by hemispherical characteristic functions.  This is not just a
formal observation.  It tells us what KM theory can and cannot
represent, why multilayer composition preserves angular rank, and where
the boundary lies between media for which the two-flux description is
adequate and those for which it is not.

Throughout, we work in the steady-state, plane-parallel,
azimuthally-symmetric setting with no emission.

\section{The radiative transfer equation}

The starting point is the steady-state, plane-parallel radiative
transfer equation~\cite{Chandrasekhar1960} for the specific intensity $I(z,\mu)$:
\begin{equation}\label{eq:rte}
  \mu\,\frac{\partial I}{\partial z}
  = -(\sigma_a+\sigma_s)\,I
    + \frac{\sigma_s}{2} \int_{-1}^{1} p(\mu,\mu')\,I(z,\mu')\,\dd\mu',
\end{equation}
where $z$ is depth along the slab normal, $\mu=\cos\theta$ is the
direction cosine, $\sigma_a$ and $\sigma_s$ are the absorption and
scattering coefficients, and $p(\mu,\mu')$ is the phase function,
normalized so that $\int_{-1}^{1}p(\mu,\mu')\,\dd\mu'=2$ for
each~$\mu$.  (Some authors normalize to unity or to~$4\pi$; only
numerical prefactors change.)

The key feature of this equation is that it is infinite-dimensional in
angle.  At every depth~$z$, the unknown is not a number or a pair of
numbers but a function on~$[-1,1]$.  Any practical method of solution
must reduce this to something finite.  Different reduction
strategies---spherical harmonics, discrete ordinates, Monte Carlo,
two-flux averaging---define different families of approximate transport
theories.  The question is where KM theory sits in this landscape.

\section{Functional setting}

We need a small amount of notation.  Let
$\mathcal{H}=L^2([-1,1])$ be the space of square-integrable functions
on the angular domain, with the standard inner product
$\langle f,g\rangle = \int_{-1}^{1}f(\mu)\,g(\mu)\,\dd\mu$.

The transport operator acts on $I(z,\cdot)\in\mathcal{H}$ as
\begin{equation}\label{eq:transport-op}
  \mathcal{T}[I]
  = \mu\,\partial_z I + (\sigma_a+\sigma_s)\,I - \frac{\sigma_s}{2}\,\mathcal{K}_p[I],
\end{equation}
where $\mathcal{K}_p$ is the integral scattering operator:
$(\mathcal{K}_p[I])(\mu)=\int_{-1}^{1}p(\mu,\mu')\,I(\mu')\,\dd\mu'$.
Since $\mathcal{H}$ is infinite-dimensional, so is the state space of
the RTE\@.  Any finite-dimensional approximation amounts to choosing a
subspace $V\subset\mathcal{H}$ and projecting the physics onto it.  The
character of the approximation---what it captures and what it loses---is
determined entirely by the choice of~$V$.

\smallskip
\noindent\textit{Remark on rigor.}\quad
We treat $\mathcal{T}$ formally as an unbounded operator on
$\mathcal{H}$; all manipulations below (projections, inner products,
term-by-term integration) can be justified for smooth, compactly
supported test functions and extended by density to $L^2$.  We will not
belabor domain questions, since the projection itself maps to a
finite-dimensional subspace on which all operators are bounded.

\begin{table}[t]
\centering
\caption{Summary of notation for hemispherical phase-function integrals
and derived quantities.}
\label{tab:notation}
\smallskip
\begin{tabular}{@{}ll@{}}
\toprule
Symbol & Meaning \\
\midrule
$\chi^+$, $\chi^-$ & Forward/backward hemispherical indicator functions \\
$\mathcal{P}$ & Hemispherical projection operator \\
$\bar{p}_{++}$ & $\int_0^1\!\int_0^1 p(\mu,\mu')\,\dd\mu'\,\dd\mu$
  \quad (forward-to-forward) \\
$\bar{p}_{-+}$ & $\int_0^1\!\int_{-1}^0 p(\mu,\mu')\,\dd\mu'\,\dd\mu$
  \quad (backscatter fraction) \\
$K = 2\sigma_a$ & KM absorption coefficient \\
$S = \sigma_s\bar{p}_{-+}$ & KM scattering coefficient \\
$g = \frac{1}{2}\int_{-1}^{1}\mu\,p(\mu)\,\dd\mu$
  & Asymmetry parameter (invisible to $\mathcal{P}$) \\
$\varepsilon = \|I^{(\perp)}\|/\|I\|$
  & Relative projection error \\
$\tau^* = \tau(1-g)$ & Reduced optical thickness \\
$\gamma = \sqrt{K(K+2S)}$ & KM eigenvalue parameter \\
\bottomrule
\end{tabular}
\end{table}

\section{Comparison: the diffusion ($P_1$) approximation}

Before getting to KM, it helps to recall the diffusion approximation,
because it is also two-dimensional and often confused with KM theory.
As we will show in Section~\ref{sec:central}, KM theory projects onto
the hemispherical subspace $\mathrm{span}\{\chi^+,\chi^-\}$; diffusion
theory projects onto a different two-dimensional subspace.

The $P_1$ approximation expands the intensity in the first two Legendre
polynomials:
\begin{equation}\label{eq:P1}
  I(z,\mu) \approx \tfrac{1}{2}\,\phi(z) + \tfrac{3}{2}\,\mu\,J(z),
\end{equation}
where $\phi$ is the scalar fluence and $J$ is the net current.
Scattering anisotropy enters through the reduced
scattering coefficient $\sigma_s^*=\sigma_s(1-g)$, where
$g=\frac{1}{2}\int_{-1}^{1}\mu\,p(\mu)\,\dd\mu$ is the asymmetry parameter.
The underlying subspace is
$V_{P_1}=\mathrm{span}\{1,\mu\}$: diffusion theory retains the first
angular moment of the phase function and captures angular gradient
structure through the $\mu$ component.

KM theory, as we will see, uses a different two-dimensional subspace
that captures angular \emph{sign} structure---forward versus
backward---but no gradient.  Same rank, completely different physics.
The two approximations are complementary, not equivalent.  Consider
Henyey--Greenstein scattering with $g=0.95$: diffusion theory sees~$g$
through $\sigma_s^*=\sigma_s(1-g)$ and correctly identifies the slab as
transport-thin.  KM theory sees only $\bar{p}_{-+}$ and cannot
distinguish $g=0.90$ from $g=0.99$ when their backscatter fractions are
similar---precisely the regime where the distinction matters most.

\section{The hemispherical projection}

Now we get to the heart of the matter.  Define two hemispherical
indicator functions:
\begin{equation}\label{eq:chi}
  \chi^+(\mu) = \begin{cases} 1, & \mu>0,\\ 0, & \mu\le 0,\end{cases}
  \qquad
  \chi^-(\mu) = \begin{cases} 0, & \mu\ge 0,\\ 1, & \mu<0.\end{cases}
\end{equation}
These are orthogonal ($\langle\chi^+,\chi^-\rangle=0$) and normalized
($\langle\chi^+,\chi^+\rangle=\langle\chi^-,\chi^-\rangle=1$).

The projection operator $\mathcal{P}$ maps any angular distribution to
its hemispherical averages:
\begin{equation}\label{eq:proj}
  \mathcal{P}f = \langle f,\chi^+\rangle\,\chi^+
                + \langle f,\chi^-\rangle\,\chi^-.
\end{equation}
In plain language: replace $f(\mu)$ for all $\mu>0$ by its average over
the forward hemisphere, and replace $f(\mu)$ for all $\mu<0$ by its
average over the backward hemisphere.

Three properties matter.  First, $\mathcal{P}^2=\mathcal{P}$
(idempotence)---projecting twice is the same as projecting once.
Second, $\mathcal{P}$ is self-adjoint with respect to the flat $L^2$
inner product $\langle f,g\rangle = \int f\,g\,\dd\mu$.
Third, the range of
$\mathcal{P}$ is two-dimensional:
$\range(\mathcal{P})=\mathrm{span}\{\chi^+,\chi^-\}$.

This is the coarsest angular discretization that preserves the
distinction between forward and backward.  It respects the slab
geometry---light going left versus light going right---but resolves
nothing finer than that within either hemisphere.  That is what gives KM
theory its characteristic combination of geometric realism and angular
coarseness.

\smallskip
\noindent\textit{Remark on the inner product.}\quad
The flat $L^2$ measure $\langle f,g\rangle = \int f\,g\,\dd\mu$ is the
natural choice for projecting the integral scattering operator, since all
directions within a hemisphere contribute equally to the hemispherical
flux.  Under the transport-weighted measure
$\langle f,g\rangle_\mu = \int|\mu|\,f\,g\,\dd\mu$---which makes the
streaming operator self-adjoint---the hemispherical basis functions are
not orthogonal, and $\mathcal{P}$ would not be an orthogonal projection.
The flat-measure projection minimizes the $L^2([-1,1])$ norm of the
residual, not the flux-weighted norm; the two criteria differ most for
grazing angles ($\mu\to 0$), which contribute little to net flux but
equally to the flat-measure error.  In practice, the distinction is
secondary for optically thick media where the angular distribution is
already nearly flat within each hemisphere.

\section{The central result: KM as projected transport}\label{sec:central}

\medskip
\noindent\fbox{\parbox{0.96\textwidth}{%
\textbf{Main result.}\quad
The Kubelka--Munk transport operator is exactly the rank-2 Galerkin
projection of the radiative transfer operator onto the hemispherical
subspace $\mathrm{span}\{\chi^+,\chi^-\}$:
\[
  \mathcal{T}_{\mathrm{KM}} = \mathcal{P}\,\mathcal{T}\,\mathcal{P}.
\]
The KM coefficients are $K=2\sigma_a$ and $S=\sigma_s\bar{p}_{-+}$,
where $\bar{p}_{-+}$ is the hemispherical backscatter fraction of the
phase function.  The projection is idempotent ($\mathcal{P}^2=\mathcal{P}$),
its kernel is infinite-dimensional, and its rank is preserved under
multilayer composition.}}
\medskip

The claim is that the KM operator is exactly the ``sandwich''
\begin{equation}\label{eq:sandwich}
  \mathcal{T}_{\mathrm{KM}} = \mathcal{P}\,\mathcal{T}\,\mathcal{P}.
\end{equation}
That is: restrict the intensity to the hemispherical subspace, apply the
full transport operator, then project back.

We now carry out this projection in full.  Write the projected intensity
as
\begin{equation}\label{eq:proj-I}
  (\mathcal{P}I)(z,\mu) = I^+(z)\,\chi^+(\mu) + I^-(z)\,\chi^-(\mu),
\end{equation}
where the hemispherical fluxes are
\[
  I^+ = \int_0^1 I\,\dd\mu, \qquad I^- = \int_{-1}^0 I\,\dd\mu.
\]

\subsection*{Step 1: The streaming term}

Apply the streaming part of $\mathcal{T}$ to the projected intensity and
take the inner product with $\chi^+$:
\begin{equation}\label{eq:stream}
  \bigl\langle \mu\,\partial_z(\mathcal{P}I),\,\chi^+\bigr\rangle
  = \frac{\dd I^+}{\dd z}\int_0^1 \mu\,\dd\mu
  = \frac{1}{2}\,\frac{\dd I^+}{\dd z}.
\end{equation}
For the backward hemisphere, the same computation gives
$\langle\mu\,\partial_z(\mathcal{P}I),\chi^-\rangle
= \frac{1}{2}\,\frac{\dd I^-}{\dd z}$
(with $\int_{-1}^{0}\mu\,\dd\mu = -\frac{1}{2}$, the sign being
absorbed into the direction of $I^-$).  The factor $\frac{1}{2}$
from $\int_0^1\mu\,\dd\mu$ is the origin of the classical factor-of-2
relationship between KM and transport coefficients discussed many times
in the literature~\cite{Vargas1997}; here it emerges from the projection.

\subsection*{Step 2: The extinction term}

The extinction (absorption plus out-scattering) contribution is
\begin{equation}\label{eq:extinct}
  \bigl\langle (\sigma_a+\sigma_s)\,\mathcal{P}I,\,\chi^+\bigr\rangle
  = (\sigma_a+\sigma_s)\,I^+\,\underbrace{\langle\chi^+,\chi^+\rangle}_{=\,1}
  = (\sigma_a+\sigma_s)\,I^+.
\end{equation}

\subsection*{Step 3: The scattering integral}

This is where the phase function enters---and where it gets averaged
away.  Apply $\mathcal{K}_p$ to the projected intensity and project onto
$\chi^+$:
\begin{align}
  \bigl\langle \mathcal{K}_p[\mathcal{P}I],\,\chi^+\bigr\rangle
  &= \int_0^1 \int_{-1}^{1} p(\mu,\mu')\,
     \bigl[I^+\chi^+(\mu') + I^-\chi^-(\mu')\bigr]\,\dd\mu'\,\dd\mu
     \notag\\[4pt]
  &= I^+\!\underbrace{\int_0^1\!\!\int_0^1 p(\mu,\mu')\,\dd\mu'\,\dd\mu}
       _{\displaystyle\equiv\;\bar{p}_{++}}
    \;+\;
     I^-\!\underbrace{\int_0^1\!\!\int_{-1}^{0} p(\mu,\mu')\,\dd\mu'\,\dd\mu}
       _{\displaystyle\equiv\;\bar{p}_{-+}}.
  \label{eq:scat-proj}
\end{align}
Here $\bar{p}_{++}$ is the forward-to-forward hemispherical average of
the phase function (light staying in the forward hemisphere), and
$\bar{p}_{-+}$ is the backward-to-forward average (light scattered from
backward into forward).  The phase-function normalization
$\int_{-1}^{1}p(\mu,\mu')\,\dd\mu'=2$ for each~$\mu$ gives
\begin{equation}\label{eq:pnorm}
  \bar{p}_{++} + \bar{p}_{-+} = \int_0^1\!\int_{-1}^{1}p\,\dd\mu'\,\dd\mu = 2.
\end{equation}
By the symmetry $p(\mu,\mu')=p(-\mu,-\mu')$ of the azimuthally-averaged
phase function in slab geometry, the backward-hemisphere equation
involves the same two integrals:
$\bar{p}_{--}=\bar{p}_{++}$ and $\bar{p}_{+-}=\bar{p}_{-+}$.

\subsection*{Step 4: Assembling the KM equations}

Combining Steps 1--3 and requiring that the residual be orthogonal to
$\range(\mathcal{P})$---the Galerkin condition---we obtain for the
forward flux:
\begin{equation}\label{eq:km-fwd-deriv}
  \frac{1}{2}\,\frac{\dd I^+}{\dd z}
  = -(\sigma_a+\sigma_s)\,I^+
    + \frac{\sigma_s}{2}\bigl(\bar{p}_{++}\,I^+ + \bar{p}_{-+}\,I^-\bigr).
\end{equation}
Multiply through by~2 and collect terms using
$\bar{p}_{++}=2-\bar{p}_{-+}$ from Eq.~\eqref{eq:pnorm}:
\begin{align}
  \frac{\dd I^+}{\dd z}
  &= -2(\sigma_a+\sigma_s)\,I^+
     + \sigma_s(2-\bar{p}_{-+})\,I^+
     + \sigma_s\bar{p}_{-+}\,I^-
  \notag\\[3pt]
  &= -\bigl[2\sigma_a + 2\sigma_s
     - \sigma_s(2-\bar{p}_{-+})\bigr]\,I^+
     + \sigma_s\bar{p}_{-+}\,I^-
  \notag\\[3pt]
  &= -(2\sigma_a + \sigma_s\bar{p}_{-+})\,I^+
     + \sigma_s\bar{p}_{-+}\,I^-.
     \label{eq:collect}
\end{align}
In the last line, the $2\sigma_s$ from extinction and the $2\sigma_s$
from forward in-scattering cancel, leaving only absorption and
backscatter.
Defining the KM coefficients
\begin{equation}\label{eq:KS-def}
  \boxed{K = 2\sigma_a, \qquad S = \sigma_s\bar{p}_{-+},}
\end{equation}
where $\bar{p}_{-+}=\int_0^1\int_{-1}^{0}p(\mu,\mu')\,\dd\mu'\,\dd\mu$
is the hemispherical backscatter fraction, Eq.~\eqref{eq:collect}
becomes
\begin{equation}\label{eq:km-forward}
  \frac{\dd I^+}{\dd z} = -(K+S)\,I^+ + S\,I^-.
\end{equation}
An identical computation for the backward hemisphere (projecting onto
$\chi^-$) yields
\begin{equation}\label{eq:km-backward}
  \frac{\dd I^-}{\dd z} = (K+S)\,I^- - S\,I^+.
\end{equation}
Equations~\eqref{eq:km-forward}--\eqref{eq:km-backward} are the
Kubelka--Munk equations.  They have been derived here by pure linear
algebra---an orthogonal projection of the full transport operator onto
the hemispherical subspace $\mathrm{span}\{\chi^+,\chi^-\}$---with no
additional physical approximations beyond the projection itself.  The
choice of subspace \emph{is} the approximation; every other step is
exact.  Because
$\mathcal{P}^2=\mathcal{P}$, the operator
$\mathcal{P}\mathcal{T}\mathcal{P}$ is a genuine Galerkin projection: it
maps the two-dimensional $\range(\mathcal{P})$ to itself.  (The
projection is Galerkin in the sense that the trial and test spaces
coincide; the underlying transport operator is non-self-adjoint and no
variational principle is invoked.  The hemispherical basis functions are
piecewise constant---degree-zero discontinuous elements on $[-1,0]$ and
$[0,1]$---making this the lowest-order discontinuous Galerkin
discretization of the angular variable.)

\section{What the projection throws away}\label{sec:kernel}

The kernel of $\mathcal{P}$ consists of all functions whose
hemispherical averages vanish:
\begin{equation}\label{eq:kernel}
  \kers(\mathcal{P})
  = \Bigl\{f\in L^2([-1,1]) :
    \int_0^1 f\,\dd\mu = 0 \;\text{and}\;
    \int_{-1}^{0}f\,\dd\mu = 0\Bigr\}.
\end{equation}
This is infinite-dimensional.  It contains every angular variation
within a hemisphere: any forward-peaked structure, any oscillatory
detail, any departure from a flat distribution---as long as it averages
to zero over the hemisphere.  Two phase functions that produce the same
hemispherical backscatter fraction $\bar{p}_{-+}$ but have wildly
different angular profiles are mapped to the same KM parameters.

This is the rigorous version of what practitioners in color science and
coatings---the primary users of KM theory---have always known
intuitively: KM theory ``cannot see'' angular detail.  But the
projection framework makes it precise.  It is not that KM theory is a
poor approximation; it is an exact projection that discards a specific,
identifiable, infinite-dimensional piece of the angular distribution.
The discarded piece has a name---$\kers(\mathcal{P})$---and everything
we need to know about when KM theory will work well and when it will not
comes down to how much energy the true solution carries in that kernel.

For instance, the function $f(\mu)=\mu - \tfrac{1}{2}$ on $[0,1]$
(extended antisymmetrically to $[-1,0]$) has zero hemispherical average
on each half---it belongs to $\kers(\mathcal{P})$---yet it carries the
linear angular gradient that distinguishes a forward-peaked distribution
from an isotropic one.

\begin{figure}[t]
\centering
\includegraphics[width=\textwidth]{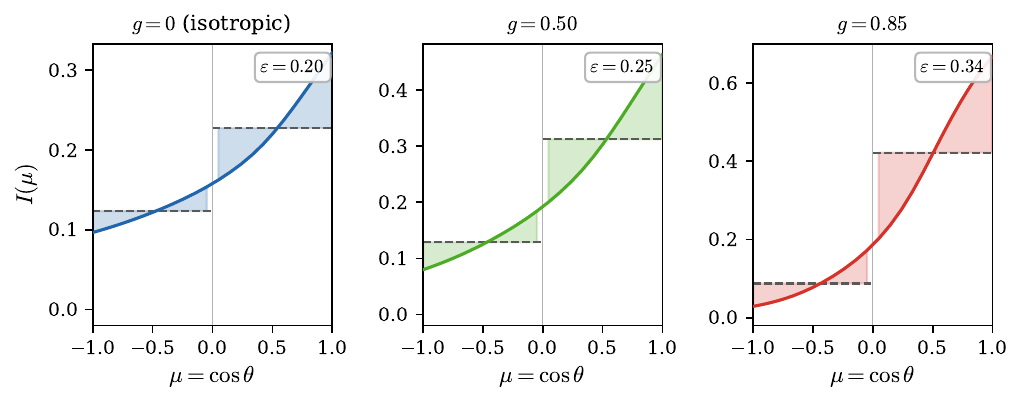}
\caption{Angular intensity $I(\mu)$ at the slab midplane
($\tau=5$, $\omega=0.9$) for three scattering regimes (32-stream
discrete-ordinates reference solution).  Dashed lines: hemispherical
projection~$\mathcal{P}I$; shaded regions: kernel component
$I^{(\perp)}=I-\mathcal{P}I$.  The projection error $\varepsilon$
grows from~0.20 (isotropic) to~0.34 ($g=0.85$).}
\label{fig:angular}
\end{figure}

\section{How this differs from discrete ordinates ($S_2$)}

The two-point discrete-ordinates method ($S_2$) also produces a pair of
coupled ODEs that look a lot like the KM equations.  It selects two
directions $\pm\mu_0$ (typically $\mu_0=1/\sqrt{3}$) and evaluates the
RTE at those specific angles.

Despite the algebraic resemblance, the methods are structurally
different.  $S_2$ is a collocation method: it demands that the equation
hold exactly at selected points and samples the phase function at
specific angle pairs.  KM theory is a Galerkin method: it integrates the
equation over hemispheres, averaging the phase function rather than
sampling it.  The quadrature weight in KM is unity over each hemisphere
(all directions contribute equally); in $S_2$ it is effectively a Dirac
delta at~$\pm\mu_0$.

The two methods respond differently to anisotropy.  $S_2$ retains
directional information at the selected ordinate; KM averages it out.
They coincide only for isotropic scattering, where the distinction
between sampling and averaging disappears.

A small numerical example makes the divergence concrete.  For the
standard slab ($\tau=5$, $\omega=0.9$) with isotropic scattering, the
$S_{32}$ reference reflectance is $R=0.518$; KM gives $0.519$ and $S_2$
gives $0.518$---all three agree.  At $g=0.50$, KM gives $R=0.434$ (an
error of~0.021 against the reference) while $S_2$ gives $R=0.591$
(error~0.179).  The $S_2$ error is nearly ten times larger, because the
forward peak evaluated at $\mu_0=1/\sqrt{3}$ distorts the point-sampled
phase function.  KM theory, by averaging over the hemisphere, is less
sensitive to the shape of the forward peak---which is both its
limitation and, in this case, its advantage.

\section{Forward-peaked scattering and the loss of~$g$}

The limitations of the hemispherical projection are most acute for
strongly forward-peaked scattering.  For Henyey--Greenstein scattering
with asymmetry parameter~$g$, the transport mean free path is
\begin{equation}\label{eq:tmfp}
  \ell^* = \frac{\ell}{1-g},
\end{equation}
where $\ell = 1/(\sigma_a+\sigma_s)$.  As $g\to 1$, the phase function
concentrates into a narrow forward peak.  The energy in this peak lies
entirely within $\kers(\mathcal{P})$: it is angular structure within the
forward hemisphere that averages to zero against the flat basis function
$\chi^+$.  The projected operator $\mathcal{P}\mathcal{T}\mathcal{P}$
sees only the total backscatter fraction and is blind to the shape of
the forward peak.

KM theory is therefore insensitive to the distinction between, say,
$g=0.9$ and $g=0.99$ whenever the backscatter fractions are similar.
All the angular structure that separates these two very different
scattering regimes resides in $\kers(\mathcal{P})$.  This insensitivity
is intrinsic to the hemispherical projection and therefore propagates to
any formalism that operates within the projected subspace, including the
compositional models of H\'ebert and Hersch~\cite{Hebert2007}.  In the
application domains where those models have been validated---printed
paper, ink-on-substrate systems---this is not a limitation in practice,
because multiple scattering in the substrate has already isotropized the
angular distribution.  The information about $g$ has been erased by the
physics, not by the model.  In forward-peaked media, however, this
information is physically present and its loss is consequential.

Corrections such as the $\delta$-Eddington~\cite{Joseph1976} and
$\delta$-$M$~\cite{Wiscombe1977} methods can
be understood precisely in this framework.  The $\delta$-$M$ method
replaces the phase function by
\[
  p(\mu,\mu') \;\longrightarrow\;
  (1-f)\,p^*(\mu,\mu') + 2f\,\delta(\mu-\mu'),
\]
where $f=g^N$ is the forward-peak fraction and $p^*$ is a smoother
residual phase function.  The $\delta$-function term, being perfectly
forward-peaked, has zero hemispherical backscatter and therefore lies
entirely in $\kers(\mathcal{P})$.  The $\delta$-$M$ truncation removes
this component \emph{before} projecting, transferring its energy into a
renormalized extinction coefficient.  In projection-theoretic language,
the modified transport problem has a solution whose kernel component
$\|I^{(\perp)}\|$ is smaller, thereby reducing $\varepsilon$.  The
procedure does not enlarge the projected subspace---KM theory still
operates in rank~2---but it ensures that the solution being projected
has less energy in the piece the projection discards.  The connection between forward-peak asymptotics and
the Cauchy kernel structure of the Henyey--Greenstein phase function is
developed in~\cite{ZellerCordery2025}: when expanded on the Motzkin
polynomial basis, the Henyey--Greenstein phase function exhibits
Cauchy/Lorentzian kernel asymptotics in the $g\to 1$ limit, providing an
independent characterization of the angular information that the
hemispherical projection discards.

\section{The thin-slab regime}

In optically thin media ($\sigma_t L\ll 1$), the intensity is dominated
by the ballistic (unscattered) component, which decays as
$\exp(-\sigma_t z/\mu)$.  This signal is sharply directional---it
remembers the angle of incidence---and a hemispherical average
misrepresents it badly.

A quantitative estimate is straightforward.  For diffuse illumination
($I=1$ for $\mu>0$ at the top surface), the ballistic contribution at
optical depth~$t$ is $I_{\mathrm{ball}}(\mu)=e^{-t/\mu}$ on the
forward hemisphere.  The hemispherical average of this signal is the
second exponential integral, $\langle I_{\mathrm{ball}},\chi^+\rangle
= E_2(t)$, and the $L^2$ norm satisfies $\|I_{\mathrm{ball}}\|^2 =
E_2(2t)$.  The projection error on the forward hemisphere alone is
therefore
\begin{equation}\label{eq:eps-thin}
  \varepsilon_{\mathrm{ball}}(t)
  = \sqrt{1 - \frac{E_2(t)^2}{E_2(2t)}}.
\end{equation}
For $t=0.5$ (the midplane of a slab with $\tau=1$), this gives
$\varepsilon_{\mathrm{ball}}\approx 0.53$; for $t=2.5$ (midplane of
$\tau=5$), $\varepsilon_{\mathrm{ball}}\approx 0.78$.  The ballistic
estimate serves as a heuristic upper bound on the actual projection
error: multiple scattering tends to isotropize the distribution and pull
it toward $\range(\mathcal{P})$, so the fully scattered solution
generally has smaller $\varepsilon$ than the unscattered one.  The
numerical solutions in
Section~\ref{sec:numerical} confirm that the true
$\varepsilon_{\mathrm{mid}}$ is considerably smaller once scattering is
included, though a rigorous proof that multiple scattering always reduces
the projection error is not attempted here.  But Eq.~\eqref{eq:eps-thin} establishes that
$\varepsilon\to O(1)$ for optically thin slabs, confirming that the
two-flux description is unreliable in that regime.

Conversely, in optically thick media ($\sigma_t L\gg 1$), multiple
scattering randomizes the angular distribution within each hemisphere.
The true intensity moves toward $\range(\mathcal{P})$, the projection
error shrinks, and KM theory becomes increasingly faithful.  This is a
general property of projection methods: they work best when the true
solution already lives near the subspace you are projecting onto.

\section{Projection error and applicability}\label{sec:proj-error}

The previous two sections point to a general principle.  At each
depth~$z$, the true intensity decomposes as
$I = \mathcal{P}I + I^{(\perp)}$, where
$I^{(\perp)}\in\kers(\mathcal{P})$ is the component the projection
throws away.  The relative projection error,
\begin{equation}\label{eq:eps}
  \varepsilon(z) = \frac{\|I^{(\perp)}(z,\cdot)\|}{\|I(z,\cdot)\|},
\end{equation}
is the natural figure of merit.  When $\varepsilon$ is small, KM theory
is accurate.  When it is not, no amount of algebraic sophistication in
the compositional framework can compensate, because the information has
already been lost at the projection step.

\medskip\noindent\textbf{Where KM works well.}\quad
Optically thick, multiply scattering media with moderate
anisotropy---paper, paint films, coatings, textile layers.  After many
scattering events the angular distribution within each hemisphere is
nearly flat, so $\|I^{(\perp)}\|$ is small and the hemispherical
averages capture almost everything.

This is the domain in which the compositional models of H\'ebert and
Hersch~\cite{Hebert2006, Hebert2007, Mazauric2014} achieve their
impressive accuracy (typical color differences $\Delta E_{94}^*<1$).
The key mechanism is that the physics of multiple scattering isotropizes
the angular distribution \emph{before} the KM projection acts on
it---the information about~$g$ and the shape of the forward peak has
already been destroyed by the medium, not by the model.  Paper
substrates are heavily scattering: by the time light has bounced through
a mat of cellulose fibers, the angular distribution within each
hemisphere is nearly flat and $\|I^{(\perp)}\|$ is genuinely small.
Ink layers attenuate primarily by Beer--Lambert absorption---there is
almost no scattering to have a phase function of.  Interfaces are
handled by Fresnel coefficients---exact geometric optics and
angular-rank-preserving by the argument of Section~\ref{sec:rank}.
The excellent accuracy reported for these systems is real and
well-earned; the projection-error framework explains \emph{why}.

Even within the paper domain, the projection is not perfect.
Edstr\"om~\cite{Edstrom2004}, comparing KM against the
discrete-ordinates solver DORT2002, found that KM reflectance errors can
reach 20\% for strongly absorbing papers.  Strong absorption reduces the
number of scattering events needed to isotropize the angular
distribution, leaving more energy in $\kers(\mathcal{P})$---exactly what
the projection-error framework predicts.

\medskip\noindent\textbf{Where KM breaks down.}\quad
Biological tissue ($g\approx 0.95$), atmospheric aerosols, sea water,
clouds---any medium where the phase function has a strong forward peak.
The peak energy lives in $\kers(\mathcal{P})$, the projection error is
large, and the KM scattering coefficient~$S$ cannot distinguish
qualitatively different scattering regimes.  This is not a deficiency of
any particular implementation; it is a structural consequence of the
rank-2 projection.  Higher-order methods---$P_N$, $S_N$, Monte
Carlo---are essential.

\medskip\noindent\textbf{Thin slabs.}\quad
The ballistic component dominates, $\varepsilon\to O(1)$, and the
measured reflectance and transmittance depend on the angular distribution
of the illumination---something the two-flux framework does not track.

\medskip
The projection-error criterion unifies these observations.  KM theory
succeeds when multiple scattering has driven the angular distribution
toward $\range(\mathcal{P})$, and it fails when significant energy
remains in $\kers(\mathcal{P})$.  The accuracy of any model that takes
KM parameters as inputs is bounded by the same criterion.

\section{Numerical illustration}\label{sec:numerical}

To make the projection-error framework concrete, we compute
$\varepsilon$ for three representative cases: isotropic scattering
($g=0$), moderate anisotropy ($g=0.50$), and strongly forward-peaked
scattering ($g=0.85$), all at total optical thickness $\tau=5$ and
single-scattering albedo $\omega=0.9$.  The boundary conditions are
diffuse illumination from above ($I(0,\mu)=1$ for $\mu>0$) and no light
from below.

The reference solution is computed by an $S_{32}$ discrete-ordinates
solver with Gauss--Legendre quadrature.  The solution is well converged:
results change by less than $10^{-4}$ in reflectance at $S_{64}$, and the 32-term Legendre
expansion of the phase function is more than adequate for $g\le 0.85$.
The KM solution is
computed directly from Eqs.~\eqref{eq:km-forward}--\eqref{eq:km-backward}
with $K$ and $S$ given by Eq.~\eqref{eq:KS-def} using the numerically
evaluated hemispherical phase-function integrals.

\begin{table}[t]
\centering
\caption{Projection error and reflectance/transmittance comparison for
three scattering regimes at $\tau=5$, $\omega=0.9$.  $R$ and $T$ are
hemispherical reflectance and transmittance; subscripts denote the
$S_{32}$ reference and KM solutions.  $\varepsilon_{\mathrm{mid}}$ is
the projection error at the slab midplane; $\varepsilon_{\max}$ is the
maximum over the slab.}
\label{tab:results}
\smallskip
\begin{tabular}{@{}ccccccccc@{}}
\toprule
$g$ & $\tau^*$ & $R_{\mathrm{SN}}$ & $R_{\mathrm{KM}}$
    & $T_{\mathrm{SN}}$ & $T_{\mathrm{KM}}$
    & $|\Delta R|$ & $\varepsilon_{\mathrm{mid}}$ & $\varepsilon_{\max}$ \\
\midrule
0.00 & 5.00 & 0.518 & 0.519 & 0.044 & 0.031 & 0.001 & 0.20 & 0.33 \\
0.50 & 2.50 & 0.413 & 0.434 & 0.099 & 0.064 & 0.021 & 0.25 & 0.35 \\
0.85 & 0.75 & 0.239 & 0.285 & 0.212 & 0.148 & 0.046 & 0.34 & 0.42 \\
\bottomrule
\end{tabular}
\end{table}

\begin{figure}[t]
\centering
\includegraphics[width=0.65\textwidth]{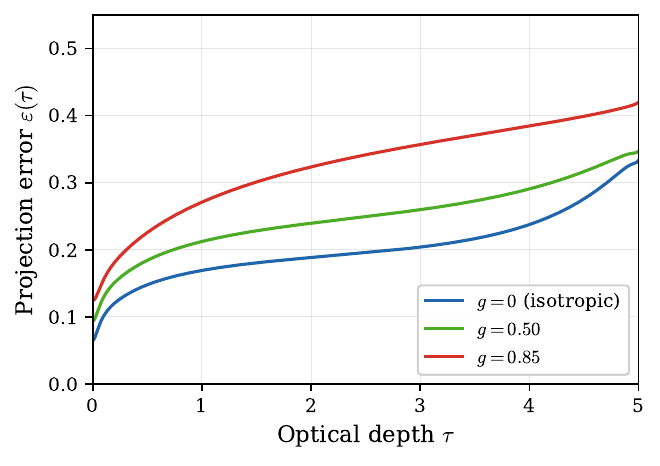}
\caption{Projection error $\varepsilon(\tau)$ through the
slab ($\tau=5$, $\omega=0.9$) for three values of~$g$.  The error is
smallest at the illuminated face and grows toward the exit; higher $g$
gives higher~$\varepsilon$ at every depth.}
\label{fig:eps}
\end{figure}

Table~\ref{tab:results} and Fig.~\ref{fig:eps} display the results.
The three cases share the same total optical depth~$\tau=5$, but the
\emph{reduced} optical depths $\tau^*=\tau(1-g)$ range from 5.0
(isotropic) to 0.75 ($g=0.85$).  In transport terms, the $g=0.85$ slab
is optically thin despite having the same $\tau$---and the projection
error reflects this.  Thus $\tau^*$ is the natural dimensionless control
parameter governing KM accuracy: the two-flux description is reliable
when $\tau^*\gg 1$, even if $\tau$ itself is modest.

\begin{figure}[t]
\centering
\includegraphics[width=0.65\textwidth]{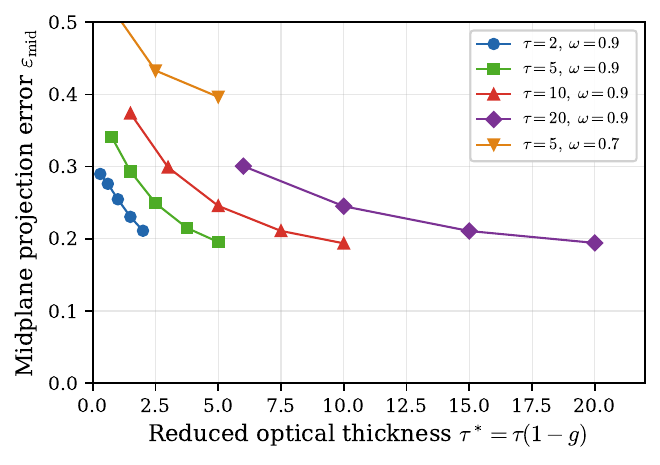}
\caption{Midplane projection error $\varepsilon_{\mathrm{mid}}$ versus
reduced optical thickness $\tau^*=\tau(1-g)$ for a range of slab
parameters.  Within each albedo family, $\tau^*$ organizes the data:
slabs with $\tau^*\gg 1$ are well described by KM theory regardless of
the individual values of~$\tau$ and~$g$.}
\label{fig:collapse}
\end{figure}

To test this more broadly, Fig.~\ref{fig:collapse} shows
$\varepsilon_{\mathrm{mid}}$ as a function of~$\tau^*$ for a range of
$\tau$ (2, 5, 10, 20), $g$ (0--0.85), and two values of~$\omega$
(0.7 and~0.9).  The curves do not collapse perfectly---$\omega$ shifts
the baseline, since lower albedo means fewer scattering events and
therefore less isotropization---but within each albedo family, $\tau^*$
organizes the data far better than either $\tau$ or~$g$ alone.  All
points with $\tau^*>5$ have $\varepsilon_{\mathrm{mid}}<0.25$; all
points with $\tau^*<1$ exceed~0.30.  This confirms that $\tau^*$ is the
right control parameter, with $\omega$ as a secondary factor.

Two features are noteworthy.  First, the midplane projection error
$\varepsilon_{\mathrm{mid}}$ tracks $g$ monotonically: 0.20 for
isotropic, 0.25 for $g=0.50$, and 0.34 for $g=0.85$.  This is exactly
what the kernel analysis of Section~\ref{sec:kernel} predicts: more
forward-peaked scattering means more energy in $\kers(\mathcal{P})$.

Second, the reflectance error $|\Delta R|$ is much smaller than
$\varepsilon$ in every case.  For isotropic scattering,
$\varepsilon_{\mathrm{mid}}=0.20$ but $|\Delta R|=0.001$.  This is
because hemispherical reflectance is itself a projected quantity---it is
$\langle I^-,\chi^-\rangle$ evaluated at the boundary---and is therefore
partially buffered from the angular errors that $\varepsilon$ measures.
The projection error bounds the observable error but does not saturate
it.  This is the numerical confirmation of the mechanism identified in
Section~\ref{sec:proj-error}: in multiply scattering media the
hemispherical observables are precisely the quantities the projection
\emph{does} capture, which is why the H\'ebert--Hersch compositional
models achieve the color accuracy they do even though the underlying
two-flux approximation is coarse.

\section{Multilayer transfer matrices}

For a homogeneous layer of thickness~$d$ with KM parameters $(K,S)$,
define
\begin{equation}\label{eq:gamma}
  \gamma = \sqrt{K(K+2S)}.
\end{equation}
The KM equations~\eqref{eq:km-forward}--\eqref{eq:km-backward} have a
matrix solution:
\begin{equation}\label{eq:matrix-sol}
  \begin{pmatrix} I^+(d) \\ I^-(d) \end{pmatrix}
  = M(d)\,\begin{pmatrix} I^+(0) \\ I^-(0) \end{pmatrix},
\end{equation}
where
\begin{equation}\label{eq:M}
  M(d) = \begin{pmatrix}
    \cosh(\gamma d) + \dfrac{K+S}{\gamma}\sinh(\gamma d)
    & \dfrac{S}{\gamma}\sinh(\gamma d) \\[8pt]
    -\dfrac{S}{\gamma}\sinh(\gamma d)
    & \cosh(\gamma d) - \dfrac{K+S}{\gamma}\sinh(\gamma d)
  \end{pmatrix}.
\end{equation}
Stacking $n$ layers with parameters $(K_i,S_i,d_i)$ gives
\begin{equation}\label{eq:stack}
  M_{\mathrm{total}} = M_n(d_n)\cdots M_2(d_2)\,M_1(d_1).
\end{equation}
Each factor is $2\times 2$, and so is the product.  The entire
multilayer calculation lives within the two-dimensional projected space.

This matrix-multiplicative structure is the algebraic machinery
exploited in the compositional models of H\'ebert and
Hersch~\cite{Hebert2007, Mazauric2014}, who showed that layer
composition via transfer matrices unifies Kubelka's layering formulae,
the Saunderson correction, and the Williams--Clapper model.  Their
approach works directly with the layer matrices, composing them without
needing to revisit the transport equation for each layer.  The present
analysis reveals the operator-theoretic origin of these matrices: they
are the natural representation of the transport operator restricted to
$\range(\mathcal{P})$.  This is why the formalism works so cleanly---it
is not an approximation to the transfer matrix; it is the transfer
matrix of the projected problem.

\section{Rank preservation}\label{sec:rank}

Since $\range(\mathcal{P})$ is two-dimensional, every projected layer
operator maps this subspace to itself.  Composing $n$ such operators
gives a map from a two-dimensional space to a two-dimensional space,
which has rank at most~2 regardless of~$n$.  In the matrix language:
every $M_i$ is $2\times 2$, and so is their product.

\begin{proposition}[Rank preservation]
Let $\mathcal{P}$ be the hemispherical projection of rank~$2$.  For any
sequence of projected layer operators $M_1,\ldots,M_n$ acting on
$\range(\mathcal{P})$, the composite operator
$M_{\mathrm{total}}=M_n\cdots M_1$ has rank at most~$2$.  No multilayer
composition can recover angular information annihilated
by~$\mathcal{P}$.
\end{proposition}

The physical consequence is important and sometimes overlooked: stacking
many KM layers does not improve the angular resolution.  A 100-layer
calculation has the same angular resolution---none within each
hemisphere---as a single-layer calculation.  More layers give you more
spectral or spatial detail, but the same two-dimensional angular view.

Improving angular resolution requires enlarging the projected subspace,
i.e., going to a higher-order method.  The four-flux matrix models
developed by Simonot et al.~\cite{Simonot2016} and the multiflux
extensions described by H\'ebert~\cite{Hebert2015} do exactly this:
they enlarge the subspace to dimension~4 (separating collimated from
diffuse components in each hemisphere) while preserving the
transfer-matrix formalism.  In the language of this paper, they replace
$\mathcal{P}$ with a higher-rank projector.

\smallskip
\noindent\textit{Remark on interfaces.}\quad
Fresnel boundary conditions do not increase the angular rank.  A planar
dielectric interface reflects a fraction~$r$ of the backward flux and
transmits~$(1-r)$ of the forward flux; in the hemispherical framework
this amounts to multiplying $I^+$ and~$I^-$ by scalar Fresnel
coefficients (averaged over the hemisphere).  The interface matrix is
therefore $2\times 2$ and diagonal, and its product with any KM layer
matrix remains $2\times 2$.  Interfaces redistribute energy between the
two hemispherical channels but do not create angular structure within
either hemisphere.  This is why the Saunderson correction and the
Williams--Clapper interface model can be absorbed into the
transfer-matrix formalism~\cite{Hebert2007} without changing its rank.

\section{Eigenstructure}

The eigenvalues of the single-layer transfer matrix are
$\lambda_\pm = e^{\pm\gamma d}$, where $\gamma=\sqrt{K(K+2S)}$.  The
spectral structure is two-dimensional regardless of layer thickness or
the complexity of the phase function from which $K$ and $S$ were
derived.

The two eigenmodes correspond to exponential growth and decay of the
flux imbalance through the slab.  These completely characterize
transport within the KM framework.  Higher angular modes---which would
be needed to describe, for example, the angular profile of the
transmitted beam---have been projected out and are inaccessible.

In the conservative limit ($K\to 0$, purely scattering medium),
$\gamma\to 0$ and the transfer matrix becomes linear in~$d$ rather than
exponential: $M(d)\to I + d\,A$, where $A$ is the generator.  The
eigenvalues $\lambda_\pm = e^{\pm\gamma d}$ coalesce at unity, producing
a degenerate case in which the flux imbalance propagates algebraically
rather than exponentially.  This is the well-known conservative KM limit
in which reflectance grows linearly with optical thickness.

\section{The inverse problem}

For a composite medium of $n$ layers, the parameter vector is
$\Theta=(K_1,S_1,\ldots,K_n,S_n)\in\bR^{2n}$.  But the measurable
quantities---total reflectance~$R$ and total transmittance~$T$---give
only two constraints.

For a single layer with known thickness, this is exactly determined, and
the classical KM inversion works.  For two or more layers, the system is
underdetermined: $2n$ unknowns, 2 equations.  You need additional
measurements---individual layer properties, spectral data at multiple
wavelengths, or angular-resolved measurements---to make the inverse
problem well-posed.

The rank-2 limitation is structural, not a matter of measurement
precision.  It is a direct consequence of the projection: since the
observables $R$ and~$T$ are hemispherical integrals---inner products
with $\chi^\pm$---they live in the two-dimensional
$\range(\mathcal{P})$ and can determine at most two independent
parameters per measurement.  Put differently, $\kers(\mathcal{P})$ is
fundamentally unobservable to any detector that measures only
hemispherical totals.  The kernel is not hidden by noise or limited
dynamic range; it is orthogonal to the measurement functionals and
produces exactly zero signal in $R$ and~$T$.

No strategy based solely on hemispherical
reflectance and transmittance can resolve more than this.  Breaking the
underdeterminacy requires either reaching into $\kers(\mathcal{P})$
through angular-resolved detection (which directly probes the discarded
angular structure) or increasing the number of projected constraints
through spectral multiplexing---measuring $R(\lambda)$ and $T(\lambda)$
at multiple wavelengths under the assumption that $K(\lambda)$ and
$S(\lambda)$ vary in a constrained way.  Spectral multiplexing does not
access $\kers(\mathcal{P})$; it provides additional hemispherical
measurements that help resolve $K$ and~$S$ as functions of wavelength,
but each measurement still lives in $\range(\mathcal{P})$.  The kernel
is infinite-dimensional, so there is a great deal of angular information
that hemispherical measurements simply cannot access.

\section{Classification and conclusions}

Multilayer KM theory is a rank-2 orthogonal projection of the radiative
transfer equation onto the hemispherical subspace
$\mathrm{span}\{\chi^+,\chi^-\}$, summarized by
$\mathcal{T}_{\mathrm{KM}}=\mathcal{P}\mathcal{T}\mathcal{P}$.  This
places it in a hierarchy of angular approximations to the RTE:

\medskip
\begin{center}
\begin{tabular}{@{}llcl@{}}
\toprule
Method & Subspace & Rank & Type \\
\midrule
Kubelka--Munk & $\{\chi^+,\chi^-\}$ & 2 & Galerkin \\
Diffusion ($P_1$) & $\{1,\mu\}$ & 2 & Galerkin \\
$S_N$ discrete ordinates & $\{\delta(\mu-\mu_i)\}$ & $N$ & Collocation \\
$P_N$ spherical harmonics & $\{P_0,\ldots,P_N\}$ & $N+1$ & Galerkin \\
Full RTE & $L^2([-1,1])$ & $\infty$ & Exact \\
\bottomrule
\end{tabular}
\end{center}

\medskip
KM theory sits at the bottom of the Galerkin ladder.  It gets the
geometry right---slab structure, layer composition, flux
conservation---without resolving any angular detail within either
hemisphere.  The title phrase ``geometric realism without angular
resolution'' is meant to capture this.  The four-flux and multiflux
models of H\'ebert et al.~\cite{Hebert2015, Simonot2016} occupy the
next rungs, distinguishing collimated from diffuse light and thereby
handling interface effects (Fresnel reflections, Saunderson corrections)
that the pure two-flux model cannot represent internally.

The kernel $\kers(\mathcal{P})$ tells us exactly what is lost: all
angular structure beyond the hemispherical average.  Rank preservation
guarantees this cannot be recovered by stacking layers.  Non-injectivity
explains why infinitely many microscopically different media can share
the same KM parameters.

The numerical illustration of Section~\ref{sec:numerical} anchors these
abstract statements in computed quantities: the projection error
$\varepsilon$ increases from 0.20 to 0.34 as the asymmetry parameter
increases from 0 to~0.85, while the reflectance error---being itself a
hemispherical quantity---remains much smaller.  The parametric sweep of
Fig.~\ref{fig:collapse} confirms that the reduced optical
thickness $\tau^*=\tau(1-g)$ governs KM accuracy across a wide range
of slab parameters, with single-scattering albedo as a secondary
factor.  The thin-slab asymptotic (Eq.~\ref{eq:eps-thin}) provides an
analytical bound showing that $\varepsilon\to O(1)$ when the ballistic
component dominates.

KM theory is therefore best understood as a well-defined low-rank
projection whose limitations are intrinsic, identifiable, and
quantifiable.  Understanding the projection lets practitioners use KM
with confidence where it belongs---optically thick, multiply scattering
media---and know precisely what is needed where it does not: a
projection onto a higher-dimensional subspace.  The compositional
transfer-matrix programme of H\'ebert, Hersch, Simonot, and
collaborators~\cite{Hebert2007, Hebert2015, Simonot2016} provides the
practical machinery for composing layers and interfaces at any flux
order, and its demonstrated accuracy in printed media was, at first
glance, surprisingly good for a rank-2 model.  The projection-error
framework explains, retroactively, why: in those systems the physics of
multiple scattering has already isotropized the angular distribution, so
the projection discards information that the medium itself has destroyed.
The present paper provides the complementary theoretical
foundation: the Galerkin projection identifies $K$ and $S$ as
hemispherical moments of the transport operator, shows that $g$ and all
higher angular structure are annihilated by the projection, explains why
the transfer-matrix formalism is natural for the projected problem, and
shows where the two-flux description stops working.

\section*{Acknowledgments}

The author thanks Bob Cordery for ongoing discussions on radiative
transport and light scattering, including the joint work on Motzkin
polynomial representations~\cite{ZellerCordery2025}, that motivated much
of this work.

The author used Claude (Anthropic) as an AI assistant for mathematical
checking, numerical framing, and manuscript preparation. The core
scientific ideas, figures, and computations are the author's own. The
author takes full responsibility for all results.



\begin{thebibliography}{99}

\bibitem{KM1931}
P.~Kubelka and F.~Munk, ``Ein Beitrag zur Optik der Farbanstriche,''
Z.\ Tech.\ Phys.\ \textbf{12}, 593--601 (1931).

\bibitem{Kubelka1948}
P.~Kubelka, ``New contributions to the optics of intensely
light-scattering materials.\ Part~I,''
J.\ Opt.\ Soc.\ Am.\ \textbf{38}, 448--457 (1948).

\bibitem{ZellerCordery2025}
C.~Zeller and R.~Cordery, ``Motzkin polynomial representations of
Henyey--Greenstein scattering and Cauchy kernel asymptotics,'' arXiv
preprint (2025).

\bibitem{Hebert2006}
M.~H\'ebert and R.~D.~Hersch, ``Reflectance and transmittance model for
recto-verso halftone prints,''
J.\ Opt.\ Soc.\ Am.~A \textbf{23}, 2415--2432 (2006).

\bibitem{Hebert2007}
M.~H\'ebert, R.~D.~Hersch, and J.-M.~Becker, ``Compositional
reflectance and transmittance model for multilayer specimens,''
J.\ Opt.\ Soc.\ Am.~A \textbf{24}, 2628--2644 (2007).

\bibitem{Mazauric2014}
S.~Mazauric, M.~H\'ebert, L.~Simonot, and T.~Fournel, ``Two-flux
transfer matrix model for predicting the reflectance and transmittance
of duplex halftone prints,''
J.\ Opt.\ Soc.\ Am.~A \textbf{31}, 2775--2788 (2014).

\bibitem{Hebert2015}
M.~H\'ebert, ``Two-flux and multiflux matrix models for colored
surfaces,'' in \emph{Handbook of Digital Imaging}, M.~Kriss, ed.\
(Wiley, 2015).

\bibitem{Simonot2016}
L.~Simonot, R.~D.~Hersch, M.~H\'ebert, and S.~Mazauric, ``Multilayer
four-flux matrix model accounting for directional--diffuse light
transfers,'' Appl.\ Opt.\ \textbf{55}, 27--37 (2016).

\bibitem{Saunderson1942}
J.~L.~Saunderson, ``Calculation of the color of pigmented plastics,''
J.\ Opt.\ Soc.\ Am.\ \textbf{32}, 727--736 (1942).

\bibitem{Mudgett1971}
P.~S.~Mudgett and L.~W.~Richards, ``Multiple scattering calculations
for technology,'' Appl.\ Opt.\ \textbf{10}, 1485--1502 (1971).

\bibitem{Star1988}
W.~M.~Star, J.~P.~A.~Marijnissen, and M.~J.~C.~van~Gemert,
``Light dosimetry in optical phantoms and in tissues: I.\ Multiple flux
and transport theory,'' Phys.\ Med.\ Biol.\ \textbf{33}, 437--454
(1988).

\bibitem{vanGemert1987}
M.~J.~C.~van~Gemert and W.~M.~Star, ``Relations between the
Kubelka--Munk and the transport equation models for anisotropic
scattering,'' Lasers Life Sci.\ \textbf{1}, 287--298 (1987).

\bibitem{Vargas1997}
W.~E.~Vargas and G.~A.~Niklasson, ``Applicability conditions of the
Kubelka--Munk theory,'' Appl.\ Opt.\ \textbf{36}, 5580--5586 (1997).

\bibitem{Edstrom2004}
P.~Edstr\"om, ``Comparison of the DORT2002 radiative transfer solution
method and the Kubelka--Munk model,''
Nord.\ Pulp Pap.\ Res.\ J.\ \textbf{19}, 397--403 (2004).

\bibitem{Chandrasekhar1960}
S.~Chandrasekhar, \emph{Radiative Transfer} (Dover, New York, 1960).

\bibitem{Joseph1976}
J.~H.~Joseph, W.~J.~Wiscombe, and J.~A.~Weinman, ``The
delta-Eddington approximation for radiative flux transfer,''
J.\ Atmos.\ Sci.\ \textbf{33}, 2452--2459 (1976).

\bibitem{Wiscombe1977}
W.~J.~Wiscombe, ``The delta-$M$ method: rapid yet accurate radiative
flux calculations for strongly asymmetric phase functions,''
J.\ Atmos.\ Sci.\ \textbf{34}, 1408--1422 (1977).

\end{thebibliography}
\end{document}